%% file: 00_main.tex
\documentclass{acmart}

\usepackage{array}
\usepackage{graphicx}
\usepackage{booktabs}

\AtBeginDocument{%
  \providecommand\BibTeX{{%
    \normalfont B\kern-0.5em{\scshape i\kern-0.25em b}\kern-0.8em\TeX}}}

\setcopyright{acmcopyright}
\copyrightyear{2022}
\acmYear{2022}
\setcopyright{acmlicensed}\acmConference[FAccT '22]{2022 ACM Conference on Fairness, Accountability, and Transparency}{June 21--24, 2022}{Seoul, Republic of Korea}
\acmBooktitle{2022 ACM Conference on Fairness, Accountability, and Transparency (FAccT '22), June 21--24, 2022, Seoul, Republic of Korea}
\acmPrice{15.00}
\acmDOI{10.1145/3531146.3533135}
\acmISBN{978-1-4503-9352-2/22/06}

\begin{document}

\title[Re-imagining Interpretability and Explainability using Sensemaking Theory]{Sensible AI: Re-imagining Interpretability and Explainability using\\ Sensemaking Theory}

\author{Harmanpreet Kaur}
\email{harmank@umich.edu}
\affiliation{%
  \institution{University of Michigan}
  \city{Ann Arbor}
  \state{MI}
  \country{USA}
}

\author{Eytan Adar}
\email{eadar@umich.edu}
\affiliation{%
  \institution{University of Michigan}
  \city{Ann Arbor}
  \state{MI}
  \country{USA}
}

\author{Eric Gilbert}
\email{eegg@umich.edu}
\affiliation{%
  \institution{University of Michigan}
  \city{Ann Arbor}
  \state{MI}
  \country{USA}
}

\author{Cliff Lampe}
\email{cacl@umich.edu}
\affiliation{%
  \institution{University of Michigan}
  \city{Ann Arbor}
  \state{MI}
  \country{USA}
}

\renewcommand{\shortauthors}{Kaur et al.}

\begin{abstract}
Understanding how ML models work is a prerequisite for responsibly designing, deploying, and using ML-based systems. With interpretability approaches, ML can now offer explanations for its outputs to aid human understanding. Though these approaches rely on guidelines for how humans explain things to each other, they ultimately solve for improving the artifact---an explanation. In this paper, we propose an alternate framework for interpretability grounded in Weick's sensemaking theory, which focuses on \textit{who} the explanation is intended for. Recent work has advocated for the importance of understanding stakeholders' needs---we build on this by providing concrete properties (e.g., identity, social context, environmental cues, etc.) that shape human understanding. We use an application of sensemaking in organizations as a template for discussing design guidelines for \textit{sensible AI}, AI that factors in the nuances of human cognition when trying to explain itself.

\end{abstract}



\maketitle

\section{Introduction}
\input{01_intro}

\section{Interpretability and Explainability}
\input{02_interpretability_rw}

\section{Sensemaking}
\input{03_sensemaking}

\section{Discussion}
\input{04_discussion}

\section{Conclusion}
\input{05_conclusion}


\begin{acks}
We thank our reviewers for their helpful comments. We are also grateful to Mitchell Gordon, Stevie Chancellor, and Michael Madaio for their feedback and support. Harmanpreet Kaur was supported by the Google PhD fellowship. 
\end{acks}

\bibliographystyle{ACM-Reference-Format}
\bibliography{0x_references}



\end{document}

%% file: 01_intro.tex
With ML-based systems being deployed in the wild, it's imperative that all stakeholders of these systems have some understanding of how the underlying ML model works. From the experts who develop algorithms to practitioners who design and deploy ML-based systems, and end-users who ultimately interact with these systems---stakeholders require varying levels of understanding of ML to ensure that these systems are used responsibly. Approaches like interpretability and explainability have been proposed as a way to bridge the gap between ML models and human understanding. These include models that are inherently interpretable (e.g., decision trees~\cite{quinlan1986induction}, simple point systems~\cite{zeng2017interpretable,jung2017simple} or generalized additive models~\cite{caruana2015intelligible,hastie1990generalized}) and post-hoc explanations for the predictions made by complex models (e.g., LIME~\cite{ribeiro2016should}, SHAP~\cite{lundberg_shap}). Tools that implement interpretability and explainability approaches have also been made available for public use. In light of this, recent work in HCI has evaluated the efficacy of these tools in  helping people understand ML models. These findings suggest that ML practitioners~\cite{kaur2020interpreting} and end-users~\cite{bansal2021does,kocielnik2019will} are not always able to make accurate judgments about the model, even with the help of explanations. In fact, having access to these tools often leads to over-trust in the ML models. Ultimately, noting that interpretability and explainability are meant for the stakeholders, recent work has proposed design guidelines for explanations based on research in the social sciences about how people explain things to each other~\cite{miller2017explainable,miller2018explanation}. Taking a human-centered or a model-centered approach, this prior work seeks to answer: \textit{what are the characteristics of an explanation that can help people understand ML models?}

Let us consider a real-world setting. Imagine you are a doctor in a healthcare organization that has decided to use an ML-based decision-support software to help with medical diagnosis. The system takes as input information about patients’ symptoms, demographics, family history, etc., and returns a predicted diagnosis. Naturally, you want to be able to overview why the software predicted a certain diagnosis before you suggest treatment based on its prediction. Further, you want to be able to explain to the patient why you (did not) trust and follow the predicted diagnosis. To aid with this, the software provider gives you access to an explanation system (e.g., LIME~\cite{ribeiro2016should}, SHAP~\cite{lundberg_shap}) which shows: (1) a local explanation (e.g., a bar chart) of the input features that were most important for the diagnosis made for a specific patient, (2) a global explanation for the features that are usually important to the model when making a prediction, and (3) an overview of each feature’s relationship with the output classes. The explanation system also includes interactive elements so you can ask ``what if'' questions based on different combinations of input features.

Is this enough to ensure that the ML-based decision-support software can be reliably used by the doctor? We claim that the answer to this question is no. This paper makes the argument that current interpretability and explainability solutions will always fall short of helping people reliably use ML-based systems for decision-making because of their focus on designing better explanations---in other words, improving an artifact. For example, while the explanation shows the symptoms that were important to the model's prediction (i.e., a local explanation), it does not tell the doctor to be cautious that the patient's other symptoms are fluctuating, that the patient belongs to a sub-group for which the model has limited training data, or that the nurses have noticed other relevant symptoms in the visiting family. From the patient's perspective, the explanation does not convey why, for example, their fear of having a particular disease (after an online symptom search or from family history) is unwarranted in this instance. These factors, that have little to do with the particular explanation, can alter the stakeholders' decision-making in significant ways. Here, \textit{we propose a specific theoretical framework to shift from improving the artifact (e.g., an explanation or explanation system) to understanding how humans make sense of complex, and sometimes conflicting, information.} Recent work supports this shift from \textit{what} an explanation should look like to \textit{who} it is intended for. Properties of the \textit{who} such as, prior experience with AI and ML~\cite{ehsan2021_who_xai}, attitude towards AI (e.g., algorithmic aversion~\cite{burton2020systematic,dietvorst2015algorithm}), the socio-organizational context~\cite{ehsan2021expanding}, have been observed as being critical to understanding AI and ML outputs. We extend this work by providing a framework for \textit{how} to incorporate human-centered principles to interpretability and explainability.

\begin{figure*}[t]
    \centering
    \includegraphics[width=\textwidth]{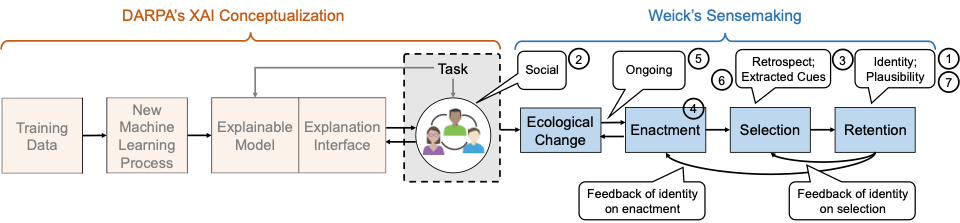}
    \vspace{-1.5pc}
    \caption{Left: DARPA's conceptualization of Explainable AI, adapted from~\cite{gunning2019darpa}. Right: Weick's sensemaking properties (1--7) categorized using the high-level Enactment-Selection-Retention organizational model, adapted from~\cite{jennings20036bconstructing}. Enactment includes properties about perceiving and acting on environmental changes; Selection, properties related to interpreting what the changes mean; and Retention, properties that describe storing and using prior experiences~\cite{kudesia2017organizational}. Our Sensible AI framework extends the existing definition of interpretability and explainability to include Weick's sensemaking properties.}
    \vspace{-1pc}
    \label{fig:sensemaking}
\end{figure*}

In this paper, we present Weick's sensemaking as a framework for envisioning the needs of people in the human-machine context. Weick describes sensemaking as, quite literally, ``the making of sense,'' or ``a developing set of ideas with explanatory possibilities''~\cite{weick1995sensemaking}. Although Weick's definition is similar to that of prior work in HCI and information retrieval, the two deviate in their goals; the latter defines sensemaking as finding representations that enable information foraging and question-answering~\cite{pirolli2005sensemaking,russell1993cost}. Weick's sensemaking is more procedural: ``placement of items into frameworks, comprehending, redressing surprise, constructing meaning, interacting in pursuit of mutual understanding, and patterning''~\cite[p.6]{weick1995sensemaking}. These processes are influenced by one's identity, environment, social, and organizational context---Weick expands these into the seven properties of sensemaking (Figure~\ref{fig:sensemaking}, Right). For example, for the doctor trying to diagnose a patient with the help of an ML-based system (with explanations), their understanding of the predicted diagnosis can be influenced by questions such as, have they recently diagnosed another patient with similar symptoms; is the patient's care team in agreement on a diagnosis; is the predicted diagnosis plausible; and, which symptoms are more visible and does the explanation present these as important to the prediction. The seven properties of sensemaking are a framework for identifying and understanding these contextual factors.

What does this knowledge of sensemaking offer to interpretability and explainability researchers and tool designers? A sensemaking perspective tells us how things beyond the individual (i.e., the environmental, social, and organizational contexts) shape individual cognition. It gives us a path forward. Prior work in organizational studies has used sensemaking to identify ways in which teams and organizations can be made more reliable. These high-reliability organizations (HROs) can serve as a template for designing \textit{Sensible AI}, AI that accounts for the nuances of human cognition when explaining itself to people. We extend the principles that make HROs reliable (e.g., a preoccupation with failure, a sensitivity to low-level operations, a reluctance to simplify anomalous situations) as guidelines for designing Sensible AI. Within our healthcare example, Sensible AI might take the form of a system that highlights the most significant ways in which a change in input features would change the predicted diagnosis; shows cases with similar input features but different diagnosis; presents input features that were considered less important by the model; asks all members of the patient care team to review the diagnosis individually first, allowing for a diversity of opinions and discussion opportunities; and asks for further explanation for cases in which the predicted diagnosis was disregarded, to inform future test cases. Our hope is that researchers and designers can translate our Sensible AI design guidelines as technical and social checks and balances in their tools, to better support human cognition as described by sensemaking.

%% file: 02_interpretability_rw.tex
\subsection{What are interpretability and explainability?}
\textit{Interpretability} is defined from a model's perspective as the ``ability to explain or to present in understandable terms to a human''~\cite[p.2]{doshi2017towards}. It serves as a proxy for other desiderata for ML-based systems such as reliability, robustness, transferability, informativeness, etc. These properties in turn promote trustworthiness, accountability, and fair and ethical decision-making~\cite{doshi2017towards,Lipton2018}. At a high-level, interpretability approaches can be categorized into glassbox models (e.g.,~\cite{quinlan1986induction,jung2017simple,zeng2017interpretable,caruana2015intelligible,lakkaraju2016interpretable,hastie1990generalized}) or post-hoc explanations for blackbox models (e.g.,~\cite{ribeiro2016should,lundberg_shap,alvarez-melis2017causal,selvaraju2017grad,simonyan2013deep}). Instantiating these approaches into user-facing tools, static explanations output by mathematical representations of interpretability now includes interactive visuals output by explainable AI. Although similarly defined, this idea of \textit{explainability} is more human-centered and is ``associated with the notion of an explanation as an interface between humans and a decision maker that is, at the same time, both an accurate proxy of the decision maker and comprehensible to human''~\cite[p.85]{arrieta2020_xai} (Figure~\ref{fig:sensemaking}, Left). Scholars have incorporated prior work from philosophy (e.g.,~\cite{hempel1948studies,pierce1878probability,vanfraassen1988pragmatic,lipton1990contrastive,pitt1988theories,grice1975logic}), the social sciences (e.g.,~\cite{lombrozo2006structure,lombrozo2012explanation,miller2017explainable,miller2018explanation,leake1991goal,slugoski1993attribution,malle2006mind,hilton1996mental,lomborg2020decoding,nisbett1977telling}), and HCI (e.g.,~\cite{bellotti2001intelligibility,norman2014some,weld2019challenge,zhu2018explainable,gillies2016human,passi2018trust,dourish2016algorithms}) with the motivation that by translating ideas from how people explain things to each other, we can design better solutions for how ML models can be explained to people. As a result, increasingly, interpretability and explainability tools include characteristics such as interactivity~\cite{amershi2014power,hohman2019gamut}, counterfactual ``what-if'' outputs~\cite{miller2021contrastive,wachter2017counterfactual}, and modular and sequential explanations~\cite{melis2021human}.

Several comprehensive reviews (e.g., ~\cite{abdul2018trends,arrieta2020_xai,liao2020questioning,wang2019designing,zhang2022towards}) synthesize and describe design considerations for the field. Based on a review of 289 core papers and 12412 citing papers, Adbul et al. highlight the trends as (1) a move from early AI work (e.g., in Expert Systems~\cite{swartout1983xplain,davis1977production}) to FAccT-centric ways of providing explanations; and (2) addressing macroscopic societal accountability in addition to helping individual users understand ML outputs~\cite{abdul2018trends}. Arrieta et al. taxonomize 409 papers to clarify terminology (e.g., interpretability, understandability, comprehensibility, etc.); describe interpretability approaches for shallow and deep learning models; and highlight the challenges for responsible AI~\cite{arrieta2020_xai}.

\subsection{Understanding the ``who'' in interpretability and explainability}
Scholars in ML, HCI, and social science communities have advocated for the importance of understanding \textit{who} the explanations are intended for. Their work identifies principles about stakeholders that are relevant in the human-machine context. \textit{Cognitive factors} (e.g., mental models, type of reasoning) have been shown to be important. For example, accurate mental models and deliberative reasoning can help avoid ML practitioners' misuse of, and over-reliance on, interpretability outputs~\cite{kaur2020interpreting}. This also applies to end-users without ML expertise~\cite{bucinca2021trust}, otherwise explanations increase the likelihood that an end-user will accept an AI’s output, regardless of its correctness~\cite{bansal2021does}. For end-users, completeness (rather than soundness) of explanations helps people form accurate mental models~\cite{kulesza2013too}. Accuracy and example-based explanations can similarly shape people's mental models and expectations, albeit in different ways~\cite{kocielnik2019will}.

\textit{Prior experience and background in ML} is also important. Variance in these can result in preset expectations, which can lead to over- or under-use of explanations~\cite{ehsan2021_who_xai}. \textit{Job- and task-dependent information needs} also shape how (much) people internalize explanations. Explanation interfaces that are interactive and collaborative can improve overall accuracy~\cite{stumpf2009interacting}. Additionally, explanations from glassbox models with fewer number of features are easier for end-users to understand~\cite{poursabzi2021manipulating}. For ML practitioners, specific types of visuals of explanations (e.g., local vs. global, sequential vs. collective) differ in how much they help them understand and debug models, and explain them to customers~\cite{hohman2019gamut,melis2021human}. Finally, \textit{social, organizational, and socio-organizational context} is important. For example,~\cite{hong2020human,veale2018fairness,madaio2020co,holstein2019improving,zhu2018explainable} all highlight the challenges of operating within an organization that either develops or employs an AI-based system. Stakeholders within and outside the organization can have conflicting needs from the system---technical interpretability and explainability approaches are unable to account for these. 

These studies from the ML, HCI, and social science communities have all highlighted relevant factors about the ``who'' in interpretability and explainability. Our proposed framework complements these evaluations: it unifies them based on sensemaking theory translated from organizational studies. We explain how individual, social, and organizational factors can affect the human-machine context, and provide a path forward that accounts for these \textit{who}-centered factors.

%% file: 03_sensemaking.tex
\begin{table*}
\fontsize{8}{10}\selectfont 
\centering
\resizebox{\linewidth}{!}{%
\begin{tabular}{>{\centering}m{0.121\linewidth}>{}m{0.42\linewidth}>{}m{0.46\linewidth}}
\multicolumn{1}{>{\centering}m{0.123\linewidth}}{\textbf{Property}} & \multicolumn{1}{>{\centering}m{0.408\linewidth}}{\textbf{Human-Human Context}} & \multicolumn{1}{>{\centering\arraybackslash}m{0.41\linewidth}}{\textbf{Human-Machine Context}}\\ 
\toprule
Identity Construction &
Sensemaking is a question about who I am as indicated by the discovery of how and what I think.  & 
Given multiple explanations, people will internalize the one(s) that support their identity in positive ways.                                                           \\ 
\hline
Social     & What I say and single out and conclude are determined by who socialized me and how I was socialized, and by the audience I anticipate will audit the conclusions I reach. & Differences in micro- and macro-social contexts affect the effectiveness of explanations.        \\ \hline

Retrospective  & To learn what I think, I look back over what I said earlier.    & Providing explanations before people can reflect on the model and its predictions negatively affects sensemaking. 
\\ \hline

Enactive & I create the object to be seen and inspected when I say or do something. & The order in which explanations are seen affects how people understand a model and its predictions.                          \\ \hline

Ongoing & Understanding is spread across time and competes for attention with other ongoing projects, by which time my interests may already have changed.  & 
The valence and magnitude of emotion caused by an interruption during the process of understanding explanations from interpretability tools change what is understood.  
\\ \hline

Focused on Extracted Cues & The `what' that I single out and embellish is only a small portion of the original utterance, that becomes salient because of context and personal dispositions.          & Highlighting different parts of explanations can lead to varying understanding of the underlying data and model. 
\\ \hline

Plausibility over Accuracy & I need to know enough about what I think to get on with my projects, but no more, which means sufficiency and plausibility take precedence over accuracy. & 
Given plausible explanations for a prediction, people are not inclined to search for the accurate one amongst these. \\
\bottomrule
\end{tabular}
}
\caption{An overview of the seven properties of sensemaking, their description in the human-human context, and our proposed claims for the human-machine context grounded in each property.}
\label{tab:seven_properties}
\end{table*}

Sensemaking describes a framework for the factors that influence human understanding; ``the sensemaking perspective is a frame of mind about frames of mind''~\cite[p.xii]{weick1995sensemaking}. It is most prominent in discrepant or surprising events. People try to put stimuli into frameworks, particularly when predictions or expectations break down. That is, when people come across new or unexpected information, they like to add structure to this unknown. The process by which they do this, why they do it, and how it affects them and their understanding of the world are all central to sensemaking.

Sensemaking subsumes interpretability\footnote{Although interpretability is defined as model-centric and explainability as human-centric, there is not yet consensus on how these terms are different from an implementation point of view. Since ``interpretability'' is commonly used in describing tools that output explanations, we use this term for the rest of the paper. We follow similar terminology choices with ML- (rather than AI-) based systems since interpretability is attributed to ML models.}. 
They share the same goal: understanding an outcome or experience. If an ML-based system could explain itself, we can verify if the reasoning is sound based on auxiliary criteria (e.g., safety, nondiscrimination), and determine whether the system meets other desiderata such as fairness, reliability, causality, and trust~\cite{doshi2017towards,Lipton2018}. Sensemaking includes all of this and more. Sensemaking not only considers the information being presented to the person doing the meaning-making, but also additional contextual nuances that affect whether and how this information is internalized. This includes factors such as, the enacted environment, the individual’s identity, their social and organizational networks, and prior experiences with similar information.

In the subsections that follow, we describe Weick's seven properties of sensemaking in the human-human context and translate them for the human-machine context (see Table~\ref{tab:seven_properties} for an overview). To concretize how these properties might affect stakeholders of ML-based systems, we present an example user vignette for each property. Prior work has applied similar methodology when translating theory~\cite{miles1994qualitative,alkhatib2021live}. While the examples are crafted based on popular press articles and research papers, they are not intended as being representative of these cases. We use them to highlight a sensemaking property, but we do not claim that the property has a causal relationship with the example, i.e., there could be other reasons for why the ML-based systems functioned the way that they are described in these articles.

\input{03a_identity}

\input{03b_social}
\input{03c_retrospective}
\input{03d_enactment}
\input{03e_ongoing}

\input{03f_cues}
\input{03g_plausibility}

\subsection{Summary}
When designing solutions for promoting human understanding of ML models, we must consider the nuances of human cognition in addition to the technical solutions which explain ML models. Sensemaking provides a set of properties that describe these nuances---each of these can be seen as a self-contained set of research questions and hypotheses that relates to the other six. As the human-machine examples show, sensemaking properties could explain external factors that shape the information that is ultimately internalized by people when they use interpretability tools.

%% file: 03a_identity.tex
\subsection{Grounded in Identity Construction}\label{sec:identity}
Identity is critical for AI/ML sensemaking because people only understand these systems in ways that they are congruent with their existing beliefs or update their beliefs while shedding a positive light on them. For interpretability, this suggests that, given multiple explanations, people will internalize the one(s) that support their identity in positive ways.

\subsubsection{Identity Construction in the \textbf{Human-Human} Context}
Sensemaking begins with the sensemaker. In this way, sensemaking is innately human-centered: ``how can \textit{I} know what \textit{I} think until \textit{I} see what \textit{I} say?''~\cite[p.18]{weick1995sensemaking}. It is grounded in the individual's need to have a clear sense of identity. People make sense of something to either support their existing beliefs or update them when applying their beliefs leads to a breakdown in their understanding. Weick notes five things of importance for identity and sensemaking~\cite[pp.23-24]{weick1995sensemaking}: (1) controlled, intentional sensemaking is triggered by a failure to confirm one's self; (2) sensemaking is grounded in the desire to maintain a consistent, positive self-conception; (3) people learn about their identities by projecting them into an environment---which includes their social, organizational, and cultural contexts---and observing the consequences; (4) sensemaking via identity construction is a mix of proaction and reaction; and (5) sensemaking is self-referential in that the self is what ultimately needs interpreting---what a given situation means is defined by the identity that an individual relies on while understanding it.

The relationship between identity and sensemaking is not limited to the individual sensemaker. The influence of social context can be seen in how identity is constructed. Weick describes this influence using three definitions of identity. First, Mead's claim that the mind and self are developed based on the communicative processes among people (i.e., social behaviorism). Individuals are comprised of ``a parliament of selves'' which reflect their various social contexts~\cite{mead1934mind}. Second, Knorr-Cetina's inclusion of social contexts based on the larger tapestry of social, organizational, and cultural norms, i.e., the macro-social~\cite{KnorrCetina1981micro}. Finally, Erez and Earley's three self-derived needs that shape identity, which include intrapersonal and interpersonal dynamics: (1) the need for \textit{self-enhancement}, seeking and maintaining a positive cognitive and affective state about the self; (2) the \textit{self-efficacy} motive, desire to perceive oneself as competent and efficacious; and (3) the need for \textit{self consistency}, desire to sense and experience coherence and continuity~\cite{erez1993culture}.

Sensemaking is made challenging by identity because the more identities that an individual has, the more ways they can assign meaning to something. Given the fluidity of identity construction, people have to grapple with several, sometimes contradicting, ways of understanding. Sometimes, this flexibility and adaptability in one's identity can be good. However, in most cases, this identity-based equivocality can lead to confusion, cognitive burden, and, in turn, lead people towards heuristics-based understanding~\cite{reason1990human}.

\subsubsection{Identity Construction in the \textbf{Human-Machine} Context}
Consider Platform X, a popular social media site which uses an ML model for content moderation, with two stakeholders in mind. First, Sharon, a 42 year old conservative in the U.S. who is against vaccination for COVID-19. Her recent posts include graphic descriptions and images of, what she claims, are the potential side-effects of getting vaccinated. Second, Avery, a 37 year old doctor who believes it is their responsibility to share unfiltered information about the COVID-19 pandemic. Several of their posts highlight the positives of getting vaccinated, and some of them present the rare potential side-effects that have been noted by medical professionals. For both Sharon and Avery, some posts have been removed by Platform X's content moderation model. 

Social media platforms usually offer an explanation for post removal to maintain their user base and help people share content in line with their policies. With interpretability tools, these platforms can support richer explanations. Based on the local explanation from an interpretability tool, Sharon is told that her post was removed due to its content type, the number of her previously flagged posts, her predicted political affiliation based on her posting history, and the topic being COVID-19. She might immediately latch on to the predicted political affiliation as \textit{the} explanation, and not try to understand the removal any further (i.e., sensemaking is not triggered because her identity remains intact). For Avery, who simply wants to share all relevant information given their identity as a doctor, the post removal might attack their needs for self-enhancement, self-efficacy, and self-consistency. As such, they might assume that the content type being graphic is the main reason for post removal---this would support their positive self-conception, and not require them to understand the model's reasoning any further. 

Interpretability tools are designed to present information in a context-free, unbiased way. But, people rarely internalize information in this static way. Weick argues that whether or how people internalize an explanation is dependent on their identity as an individual and as a part of their varying social contexts. 

\textit{Claim: Given multiple explanations, people will internalize the one(s) that support their identity in positive ways.}

%% file: 03b_social.tex
\subsection{Social}
AI/ML sensemaking is modified by social context because it represents the audience-oriented external factors that influence people as they try to understand the outputs of these systems. For interpretability, this suggests that explanations are internalized differently by people with different micro- and macro-social contexts.

\subsubsection{Social Elements of Sensemaking in the \textbf{Human-Human} Context}
Sensemaking describes human cognition. This might give it the appearance of being about the individual, but it is not. Weick notes the work on socially shared cognition (e.g.,~\cite{Resnick1991,levine1993social}) which shows that human cognition and social functioning are essential to each other. Specifically, an individual's conduct is dependent on their audience, whether this is an imagined, implied, or a physically present one~\cite{allport1985historical,bruns1961management}. Regarding the lack of a need for a physically present audience, recall Weick's reference to Mead's work on the individual being ``a parliament of selves''~\cite{mead1934mind} (see Section~\ref{sec:identity} for details on socially-grounded identity construction).

A focus on social aspects of sensemaking naturally implies that modes of communication (e.g., speech, discourse) and tools that support these also get attention, since these represent the ways in which social contact is mediated. Weick describes their importance on three levels, which exist beyond the individual: (1) inter-subjective, the conversations with others that can lead to alignment; (2) generic subjective, the socially-established norms when alignment has been achieved; and (3) extra-subjective, the culturally-established norms that do not necessarily require communication anymore. As we go from inter- to extra-subjective, the role of the implied and invisible audience becomes increasingly prominent. This, in turn, shapes the modes and tools of communication necessary for sensemaking.

\subsubsection{Social Elements of Sensemaking in the \textbf{Human-Machine} Context}
Consider the model developed for predicting diabetic retinopathy (DR) based on healthcare data (predominantly eye fundus photos) collected in the U.S.~\cite{beede2020human}. The U.S. healthcare system is consistent across organizations---there is low variability in how eye fundus photos are captured, how the medical records are stored, and who (a generalist or specialist doctor) makes a diagnosis. However, when the same model was applied to a different social and cultural context---in Thailand, where healthcare is dependent on individual providers and patient needs in different regions---it failed in unanticipated ways. 

First, there is the issue with the data itself. Several countries in Southeast Asia, including Thailand, do not have dedicated rooms for capturing fundus photos, making the photos inconsistent in opacity and leading to potentially inaccurate predictions. Second, there are established norms around the results of a DR screening test. While it is often expected to receive results immediately in the U.S. healthcare system, this is less common in Thailand, with fewer technicians, doctors, and specialists. Patients living in smaller towns have to travel to larger cities for appointments with specialists. A patient who is anticipating their DR result 4-5 weeks later might not have budgeted enough time for travel, based on a referral on the same day as the DR screening test visit. 

While interpretability tools may offer an explanation, these explanations are limited to the model and the training dataset. Weick's perspective suggests that it might not be enough to explain the prediction, due to the variability in people's social contexts when using predictions in real-world settings; recent work on domain and distributional shifts in ML datasets supports this perspective~\cite{koh2021wilds}. 

\textit{Claim: Differences in micro- and macro-social contexts affect the effectiveness of explanations.}

%% file: 03c_retrospective.tex
\subsection{Retrospective}\label{sec:retrospective}
Retrospection or reflective thinking influences AI/ML sensemaking by engaging people in deliberately thinking about the diverse interpretations of outputs when trying to understand these systems, instead of following the more automated, heuristics-based, reasoning pathways. For interpretability, this suggests that providing explanations before people can reflect on the model and its predictions negatively affects sensemaking. 

\subsubsection{Retrospective Sensemaking in the \textbf{Human-Human} Context}
Sensemaking is retrospective because the object of sensemaking is a \textit{lived experience}. Weick describes the retrospective nature of sensemaking as the most important, but perhaps the least noticeable, property. The reason it so frequently goes unnoticed is because of how embedded retrospection is in the sensemaking process. Retrospective sensemaking is derived from the work of Schutz, who believes that meaning is ``merely an operation of intentionality, which\ldots only becomes visible to the reflective glance''~\cite{schutz1976fragments,schutz1972phenomenology}. The lived timeframe being considered for reflection can be the short- or long-term past, ranging from minutes, days, and years to ``as I begin the latter portion of a long word, my utterance of the first part is already in the past''~\cite[p.44]{hartshorne1962}.

The retrospective process starts with an individual’s present circumstances, and those shape the past experiences selected for sensemaking. Reflection happens in the form of a cone of light that starts with the present and spreads backwards. In this way, the cues of the past lived experience that are paid attention to for sensemaking depend on how the present is shaped. The challenge lies in \textit{which} present to consider. People typically have several things on their mind at the same time, be it multiple projects at work or personal goals. With these, they have a multitude of lenses that they could apply for the reflective sensemaking process---the object of their sensemaking thus becomes equivocal. When dealing with equivocality, people are already overwhelmed with information and providing more details is often not helpful. ``Instead, they need values, priorities, and clarity about preferences to help them be clear about which projects matter''~\cite[p.28]{weick1995sensemaking}. In looking for clarity on which meaning to select, people are prone to a hindsight bias~\cite{staw1975attribution}. They select the most plausible story of causality for the outcome that they are trying to explain (Section~\ref{sec:plausibility} describes this property of sensemaking: being driven by plausibility over accuracy).

\subsubsection{Retrospective Sensemaking in the \textbf{Human-Machine} Context} 
For ML-based systems, the model and its predictions are the ``lived experiences.'' Consider a radiologist tasked with reading chest radiographs to determine if a patient has COVID-19. The hospital has purchased an ML-based image classification system. To help determine if the predictions makes sense, the software also provides saliency maps (an interpretability approach).

\begin{figure*}
    \centering
    \includegraphics[width=0.85\textwidth]{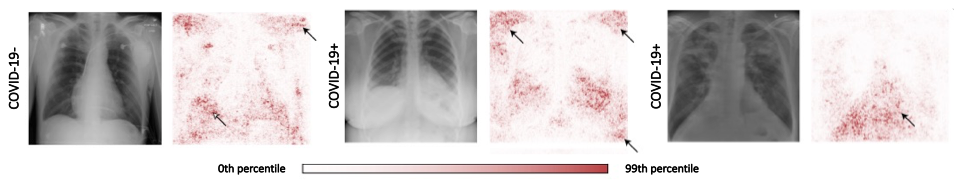}
    \vspace{-0.5pc}
    \caption{Saliency maps for chest radiographs, adapted from~\cite{degrave2021ai}.}
    \label{fig:chest_radiograph}
    \vspace{-1.25pc}
\end{figure*}

By immediately providing an explanation, the interpretability tool effectively disengages the retrospective process that helps with sensemaking. Figure~\ref{fig:chest_radiograph} shows example explanations provided to the radiologist.
As described in the accompanying research paper~\cite{degrave2021ai}, these explanations show that the ML model sometimes relies on laterality markers to make the prediction. For example, in Figure~\ref{fig:chest_radiograph}, the saliency maps highlight not only the relevant regions in the lungs as being predictive, but also some areas (see pointers) that differ based on how the radiograph was taken. These, coincidentally, are also predictive of COVID-19 positive vs. negative results, leading to a spurious correlation. 

Ideally, the radiologist evaluating the saliency map would be able to reach the same conclusion regarding these spurious correlations. However, the retrospective property would suggest that by providing this explanation without asking the radiologist to first think about what the explanation could be, the interpretability tool disengages their retrospective sensemaking process. This makes it easier for the radiologist to craft a plausible narrative that agrees with the model's prediction instead of analyzing the radiograph in detail and accurately understanding the model. When they immediately have the explanation, there is no cognitive need for the radiologist to understand the intricacies of the model, which increases the likelihood of them missing the issues with the model. Prior work on stakeholders' use of interpretability tools corroborates this perspective: people expect far more from interpretability tools than their actual capabilities and, in doing so, often end up over-trusting and misusing them~\cite{bansal2021does,kaur2020interpreting}.  

\textit{Claim: Providing explanations before people can reflect on the model and its predictions negatively affects sensemaking.}

%% file: 03d_enactment.tex
\subsection{Enactive of Sensible Environments}
Enactment is critical for AI/ML sensemaking because it represents how (much) people understand these systems---it reflects the parts of these systems that people understand, and then build on, over time. For interpretability, this suggests that the order in which explanations are seen affects how people understand a model and its predictions.

\subsubsection{Enactment in the \textbf{Human-Human} Context}
When we are tasked with making sense of something, it might appear to belong to an external environment that we must observe and understand. Weick argues that this is not the case, that sensemaking works such that ``people often produce part of the environment they face''~\cite[p.30]{weick1995sensemaking}. It is not just the person, rather, the person and their enacted environment that is the unit of analysis for sensemaking~\cite{pondy1979beyond}.

This environment that provides the necessary context for sensemaking is not a monolithic, fixed environment that exists external to people. Rather, people act, and their actions shape the environmental context needed for sensemaking: ``they act, and in doing so create the materials that become the constraints and opportunities they face.''~\cite[p.31]{weick1995sensemaking}. Here, Weick is influenced by Follett, who claims that there is no subject or object in meaning-making. There is no meaning that one understands as the ``result of the process;'' there is just a ``moment in process''~\cite[p.60]{follett1924creative}. As such, this meaning is inherently contextual in that it is shaped by the cycle of action-enaction between the human and their environment.

Weick cautions against two things with the enactive nature of sensemaking. First, to not restrict our definition of action in shaping our environment. Action here could mean creating, reflecting, or interpreting: ``the idea that action can be inhibited, abandoned, checked, or redirected, as well as expressed, suggests that there are many ways in which action can affect meaning other than by producing visible consequences in the world'' (\cite{blumer1969symbolic}, described by Weick~\cite[p.37]{weick1995sensemaking}). Second, the enacted environments do not need to embody existing ones. People want to believe that the world is defined using pre-given features, i.e., knowledge and meaning exist, we just need to find them. This is called Cartesian anxiety: ``a dilemma: either we have a fixed and stable foundation for knowledge, a point where knowledge starts, is grounded, and rests, or we cannot escape some sort of darkness, chaos, and confusion''~\cite[p.140]{varela2016embodied}. When faced with equivocal meanings, people want to select ones that reduce Cartesian anxiety. But, in doing so, they also enable existing, socially constructed meanings to shape their sensemaking. This can be helpful in providing the clarity of values needed when faced with equivocality, or it can privilege some meanings over others, depending on agency and power~\cite{ring1989formal}.

\subsubsection{Enactment in the \textbf{Human-Machine} Context}
Enactment is most apparent when ML-based systems are used in urgent or reactive situations, such as predictive policing. Consider PredPol, which uses location-based ML models that rely on connections between places and their historical crime rates to identify hot spots for police patrol~\cite{predpol_2020}. 

Say a police officer is monitoring PredPol to allocate patrol units to various neighborhoods. The model's predictions influence both the officer monitoring the software as well as those patrolling. Both will update their ``environment'' to be focused on certain neighborhoods. That is, they are primed to look for criminal activity in these neighborhoods. Additionally, when arrests are made using model predictions, they provide further evidence to the model that the patterns it has identified are accurate. 
In this way, the feedback loop causes the model to become increasingly biased~\cite{heaven2020predictive}. If the police officers were also provided an explanation for the model's predictions, the type of explanation and the order in which they are seen (e.g., global vs. local explanation first) changes the enacted environment for the officers. The sensemaking perspective offers several properties for how the environment could be shaped (e.g., people's identity, social network).

Interpretability tools offer different types of information (e.g., feature importances, partial dependency plots, data distributions), but do not impose an order on how this information is explored. End-users can take different paths to reaching conclusions about the model. Because sensemaking is sensitive to enacted environments, it is important to remember that any information or explanation about the model is not treated by people as static or isolated.

\textit{Claim: The order in which explanations are seen affects how people understand a model and its predictions.}

%% file: 03e_ongoing.tex
\subsection{Ongoing}\label{sec:ongoing}
The ongoing nature of AI/ML sensemaking highlights how interruptions and emotions can influence what is understood about these systems. For interpretability, this suggests that, if interrupted when viewing an explanation, the valence and magnitude of the resulting emotion can change what people understand about the model and its predictions.

\subsubsection{Sensemaking as an Ongoing Activity in the \textbf{Human-Human} Context}
Sensemaking never starts or stops; people are always in the middle of something. To think otherwise would suggest that people are able to chop meaningful moments from the flow of time, but that would be counter-intuitive because to determine whether something is ``meaningful'' would require sensemaking in the first place~\cite{rickman1979dilthey,dilthey1972rise}. 
Sensemaking is akin to being in situations of thrownness. Winograd and Flores describe these situations as having the following properties: (1) you cannot avoid acting; (2) you cannot step back and reflect on your actions, i.e., you have to rely on your intuitions; 
(3) the effects of action cannot be predicted; (4) you do not have a stable representation of the situation; (5) every representation is an interpretation, i.e., no objective analysis can be performed in the moment; and (6) language is action, i.e., people enact the situation via their descriptions of their environment, making it impossible to stay detached from it~\cite{winograd1986understanding}.

Emotion is embedded in sensemaking via the following process. Interruptions trigger arousal, i.e., a discharge in the autonomic nervous system, which convinces the individual that something in the environment has changed, that they must understand it and take appropriate action to get back to a state of flow~\cite{berscheid1983emotion,mandler1984mind}. The higher the arousal post-interruption, the stronger the emotional response and, in turn, the stronger the affect of emotion on sensemaking. Why does it matter if there is an emotional response during an ongoing sensemaking process? Emotions affect sensemaking in that recall and retrospect are dependent on one's mood~\cite{snyder1982moods}. Specifically, people recall events that are congruent with their current emotional valence. Of all the past events that might be relevant to sensemaking in a current situation, the ones we recall are not those that look the same, but those that feel the same.

\subsubsection{Sensemaking as an Ongoing Activity in the \textbf{Human-Machine} Context}
Consider the PredPol example again. Let's assume the arrest record shows that the likelihood of a legitimate arrest in an area predicted as a hot spot by the model is 40\%. The officer monitoring the model outputs is made aware of this number every time they log into the system. Imagine this happens one day: the patrol officers allocated to one of the hot spots make a legitimate arrest. The monitoring officer is commended for their role in anticipating the situation. This happens several times during the day. Thus, the monitoring officer associates positive feedback with arrests based on the model's predictions. When writing their report about the incidents, they use the explanations provided by the software to further justify their choices.

Next day, the patrol officers make another arrest in the same predicted hot spot. The monitoring officer is once again asked to record an explanation for selecting that area for patrol. Before they do so, they happen to look at social media and notice several posts showing outrage with regards to that arrest. This is an interruption, as described by the ongoing property of sensemaking. This time, when the monitoring officer is writing up their explanation, it could be that they mention that the model's predictions are not always right and highlight some other failure cases.

As we have noted before, information presented in explanations is rarely used in context-free settings. Despite being shown the same explanation, the monitoring officer could notice different aspects of it depending on whether they were interrupted, whether the interruption led to positive or negative emotional states, and the magnitude of those emotions. 

\textit{Claim: The valence and magnitude of the emotion caused by an interruption during the process of understanding explanations from interpretability tools change what is understood.}

%% file: 03f_cues.tex
\subsection{Focused on and by Extracted Cues}\label{sec:cues}
Extracted cues modify AI/ML sensemaking because they represent the (incomplete) bits of information that people rely on when trying to understand these systems. For interpretability, this suggests that highlighting different parts of explanations can lead to varying understanding of the underlying data and model. 

\subsubsection{Extracting Cues in the \textbf{Human-Human} Context}
Weick describes extracted cues as ``simple, familiar structures that are seeds from which people develop a larger sense of what may be occurring''~\cite[p.50]{weick1995sensemaking}. These extracted cues are important for sensemaking because they are taken as ``equivalent to the entire datum from which they come'' and in being taken as such, they ``suggest a certain consequence more obviously than it was suggested by the total datum as it originally came''~\cite[p.340]{james2007principles}. Sensemaking uses extracted cues like a partially completed sentence. The completed first half of the sentence constrains what the incomplete second half could be~\cite{shotter1983duality}.

Extracting cues involves two processes---noticing and bracketing---which are both affected by context. First, context affects which cues are extracted based on what is noticed by the sensemaker. \textit{Noticing} is an informal, even involuntary, observation of the environment that begins the process of sensemaking~\cite{starbuck1988executives}. Cues that are noticed are either novel, unusual, or unexpected, or those that we are situationally or personally primed to focus on (e.g., recently or frequently encountered cues)~\cite{taylor1991asymmetrical}. Second, context affects how the extracted (noticed) cues are interpreted. Without context, any cues that are extracted lead to equivocal meanings~\cite{leiter1980primer}. These situations of equivocality need a clarity of values instead of more information for sensemaking (Section~\ref{sec:retrospective}). Context can provide this clarity in the form of, for example, the social and cultural norms of the setting where sensemaking in happening. During the process of extracting cues, people are trying to form a cognitive reference map that presumes that there is a connection between the situation/outcome and the cue. However, important cues can be missed when people do not have any prior experience with the situation.

\subsubsection{Extracting Cues in the \textbf{Human-Machine} Context}
Consider the example where a company provides ML-based software to organizations to help them with hiring decisions. A marketing company uses this software to shortlist candidates by sending some questions in advance. The candidates answer these questions in a video format, and the ML-based software analyzes these videos and provides a hiring score along with an explanation. The kind of input data used by the model includes demographic information; prior experience from the candidate's resume; and tone of voice, perceived enthusiasm, and other emotion data coded by the software after analyzing the recorded video~\cite{kahn2021hirevue}.

Let's say that the marketing company is using this software to shortlist candidates for the position of a sales representative. The software shows that A is a better candidate than B and explains its ratings (based on local explanations from interpretability tools). The HR folks see that A’s rating is based on their facial expressions during the interview (they were smiling, not visibly nervous, and seemed enthusiastic). They consider these to be good attributes for a sales representative and hire A even though B is more qualified. Additional information about A's and B's qualifications is also noted in the local explanations but might not be the cues that are extracted or focused on in this instance.

Current interpretability tools present all types of information and let the user decide how to explore. Weick cautions against this unstructured exploration because it leads to equivocal alternatives for understanding an ML-based system. Which one of these alternatives is ultimately selected can be a reasonable, reflective process or entirely arbitrary.

\textit{Claim: Highlighting different parts of explanations can lead to varying understanding of the underlying data and model.}

%% file: 03g_plausibility.tex
\subsection{Driven by Plausibility rather than Accuracy}\label{sec:plausibility}
Recognizing that people are driven by plausibility rather than accuracy is critical for AI/ML sensemaking because we must account for people's inclination to only have a ``good enough'' understanding of these systems. For interpretability, this suggests that, given plausible explanations, people are not inclined to search for the accurate one amongst these.

\subsubsection{Plausibility over Accuracy in the \textbf{Human-Human} Context}
Weick argues that accuracy is nice but not necessary for sensemaking. Even when it is necessary, people rarely achieve it. Instead, people rely on plausible reasoning which is: (1) not necessarily correct but fits the facts, and (2) based on incomplete information~\cite{isenberg1986structure}.
When sensemaking, people can be influenced by what is ``interesting, attractive, emotionally appealing, and goal relevant''~\cite{fiske1992thinking}.

Weick notes eight reasons for why accuracy is secondary to sensemaking. Most important among these, \textit{first}, it is impossible to internalize the overwhelming amount of information available for sensemaking. To cope with this, people apply relevance filters to the information ~\cite{gigerenzer1991make,smith1991heuristics}. \textit{Second}, when people filter what they notice, this biased noticing can be good for action, though not for deliberation. But, deliberation is not the goal, it is ``futile in a changing world where perceptions, by definition, can never be accurate''~\cite[p.60]{weick1995sensemaking}. \textit{Third}, at the time of sensemaking, it is impossible to tell if the sensemaker's perceptions will be accurate. It is only in retrospect---after the sensemaker has taken action based on their understanding---that they evaluate their perceptions for accuracy.

With accuracy not being necessary for sensemaking, it is only natural to ask: what is? Weick claims that what is necessary for sensemaking is a good story, ``something that preserves plausibility and coherence, something that is reasonable and memorable, something that embodies past experiences and expectations, something that resonates with other people, something that can be constructed retrospectively but also can be used prospectively, something that captures both feeling and thought, something that allows for embellishment to fit current oddities, something that is fun to construct''~\cite[pp.60-61]{weick1995sensemaking}. Stories help with sensemaking because they are templates from previous attempts at making sense of similar situations. Overall, this property is often amplified by the others in that the plausible narratives could depend on people's identity, implied or actual audience, extracted cues, emotional state, etc.

\subsubsection{Plausibility over Accuracy in the \textbf{Human-Machine} Context}
Interpretability outputs, such as text or visual explanations, inherently present a story. As long as this explanation / story is plausible, there is no reason for an individual to evaluate it for accuracy. Consider the example with the radiologist again, where they are tasked with deciding whether a chest radiograph shows that the patient has COVID-19. Their decision-making is supported by an ML-based software that has been trained on publicly available chest radiograph datasets. To help them understand the model's reasoning for a prediction, the radiologist has access to saliency maps as interpretable outputs (Figure~\ref{fig:chest_radiograph}). 

According to Weick, when using the saliency map to determine whether the model's prediction makes sense, the radiologist is essentially searching for a plausible story that explains the prediction. The explanations in Figure~\ref{fig:chest_radiograph} show some areas inside the lungs as relevant, a plausible reason for predicting COVID-19. The radiologist could believe this plausible explanation and choose to follow it. Human evaluations of interpretability tools show that this confirmatory use of explanations is often the case, even when explanations reveal issues with the underlying model~\cite{bansal2021does,bucinca2021trust,kaur2020interpreting}.

Let's say that the radiologist was not immediately convinced that the prediction was accurate after seeing the saliency maps. Maybe they looked at one of them (e.g., Figure~\ref{fig:chest_radiograph}-Middle) and noticed that the radiograph's edges (by the person's shoulders and diaphragm) were also salient for the prediction. Even with this observation, the radiologist is looking for a plausible story. Perhaps the patient was coughing and could not stay still when the radiograph was being captured? That could explain the lateral markers for a COVID-19 positive patient. The model is relying on spurious correlations, but, with the role of plausibility in sensemaking, the radiologist might not try to accurately interpret the saliency map.

\textit{Claim: Given plausible explanations for a prediction, people are not inclined to search for the accurate one amongst these.}

%% file: 04_discussion.tex
We propose a framework for Sensible AI to account for the properties of human cognition described by sensemaking. This has the potential to refine the explanations from interpretability tools for human consumption and to better support the human-centered desiderata of ML-based systems. How do we do this? Once again, Weick (along with his colleagues) proposes a solution: to explicitly promote or amplify sensemaking, we can follow the model of \textit{mindful organizing}~\cite{weick2015managing}. Sensemaking and \textit{organizing} are inextricably intertwined. While sensemaking describes the meaning-making process of understanding, organizing describes the final outcome (e.g., a map or frame of reference) that represents the understanding. They belong to the same mutually interdependent, cyclical, recursive process---sensemaking is the process by which organizing is achieved~\cite{glynn2020organizing,weick2005organizing}. \textit{Mindfulness} is expressed by actively refining the existing categories that we use to assign meaning, and creating new categories as needed for events that have not been seen before~\cite{langer1989minding,weick2015managing,vogus2012organizational}.

Mindful organizing was proposed after observing high-reliability organizations (HROs). HROs are organizations that have successfully avoided catastrophes despite operating in high-risk environments~\cite{roberts1990some,weick1999organizing}. Examples of these include healthcare organizations, air traffic control systems, naval aircraft carriers, and nuclear power plants. Mindful organizing embodies five principles consistently observed in HROs: (1) \textbf{preoccupation with failure}, anticipating potential risks by always being on the lookout for failures, being sensitive to even the smallest ones; (2) \textbf{reluctance to simplify}, wherein each failure is treated as unique because routines, labels, and cliches can stop us from looking into details of an event; (3) \textbf{sensitivity to operations}, a heightened awareness of the state of relevant systems and processes because systems are not static or linear, and expecting uncertainty in anticipating how different systems will interact in the event of a crisis; (4) \textbf{commitment to resilience}, prioritizing training for emergency situations by incorporating diverse testing pathways and team structures, and when a failure occurs, trying to absorb strain and preserve function; and (5) \textbf{deference to expertise}, assuming that people who are in the weeds---often lower-ranking individuals---have more knowledge about the situation, and valuing their opinions. Our proposal for Sensible AI encompasses designing, deploying, and maintaining systems that are reliable by learning from properties of HROs. Table~\ref{tab:design_hros} presents the corresponding principles of HROs that serve as inspiration for each idea.

\begin{table*}[t]
  \centering
  \includegraphics[width=\textwidth]{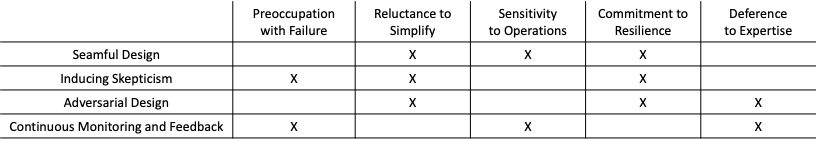}
  \caption{Principles of high-reliability organizations (columns) that inspired our design ideas (rows) for Sensible AI.}
  \label{tab:design_hros}
  \vspace{-1pc}
\end{table*}

\subsection{Seamful Design}
We can help people understand AI and ML by giving them the agency to do so. Often, ML-based systems and interpretability tools are designed with seamless interaction and effortless usability in mind. However, this can engage people's automatic reasoning mode, leading them to use ML outputs without adequate deliberation~\cite{bucinca2021trust,kaur2020interpreting,bansal2021does}. Highlighting complex details of ML outputs and processes---seamful design~\cite{inman2019beautiful}---can promote the reluctance to simplify that has helped HROs. It can also add a sensitivity to operations when changes to inputs for models can be clearly seen in the outputs. Enhancing reconfigurability of ML models and training people to understand their complexity can reduce automatic, superficial evaluations. Increasing user control in the form of seamful design has the added benefit of introducing opportunities for informational interruptions, which are helpful for the commitment to resilience seen in HROs. While current interpretability tools have interactive features that provide additional information as needed, contextualizing this information using narratives can help people maintain overall situational awareness and avoid dysfunctional momentum when using ML-based systems. For example, when a doctor is viewing a predicted diagnosis, a Sensible AI system could prompt them to view cases with similar inputs but different diagnoses. Next, we discuss ways to design these systems without overloading the end-user with features, interactivity, and information.

\subsection{Inducing Skepticism}
One way to reduce over-reliance on generalizations and known information---both common outcomes of sensemaking---is to create situations in which people would ask questions. We call this inducing skepticism, an idea suggested in prior work as a strategy for promoting reflective design~\cite{sengers2005reflective}. Inducing skepticism can foster a preoccupation with failure, an HRO principle that encourages cultivating a doubt mindset in employees. HRO employees are always on the lookout for anomalies, they interpret any new cues from their systems in failure-centric ways, and collectively promote wariness. This can be incorporated in ML-based systems, for example, by suggesting that end-users ask about how a particular prediction is unique or similar to other data points, questioning outputs of interpretability tools sometimes (e.g., ``does this feature importance value make sense?''), presenting bottom-n feature importances in an explanation instead of top-n, highlighting cases for which the model is unsure of its predictions, etc. Inducing skepticism can also be accomplished in social ways, by promoting diversity in teams, both in terms of skillsets and experience. For example, novices can prompt experts to view an AI output in more detail when they ask questions about it. This diversity is a common way in which HROs maintain their commitment to resilience. These technical and social ways of inducing skepticism have a common goal, a reluctance to simplify by adding complexity and diversity to a situation.

\subsection{Adversarial Design}
No one person can successfully anticipate all failures, even when the system induces skepticism. Adversarial design suggests relying on social and organizational networks for this task. Adversarial design is a form of political design rooted in the theory of agonism: promoting productive contestation and dissensus~\cite{disalvo2015adversarial,mouffe2013agonistics,wenman2013agonistic}. By designing Sensible AI systems with dissensus-centric features, we can increase the likelihood that \textit{someone} raises a red flag given early signals of a failure situation. 
Prior work has implemented adversarial design in the form of red teaming in technical and social ways (e.g., adversarial attacks for testing and promoting cybersecurity~\cite{abbass2011computational}, and forming teams with collective diversity and supporting deliberation~\cite{hong2004groups,hong2020contributions,gordon2022jury}, respectively). Here, HRO principles of reluctance to simplify, commitment to resilience, and deference to expertise can be observed in practice. We propose technical redundancies and social diversity to reduce unanticipated failures in understanding AI outputs, as one way of operationalizing adversarial design. Technical redundancies can be implemented as system features wherein multiple people view the same output in different contexts, giving the team a better chance of finding potential issues. Social or organizational diversity can be expanded by including people with different roles, skillsets, and opinions. The more diversity in people viewing the outputs, the higher the likelihood that they collectively discover an issue, as long as deliberation is made easy~\cite{hong2020contributions}.

\subsection{Continuous Monitoring and Feedback}
When ML-based systems are deployed in real-world settings, changes in data collection and distributional drifts are a given~\cite{koh2021wilds}. 
To manage these, researchers and practitioners have proposed MLOps---an extension of DevOps practices from software to ML-based settings---to include continuous testing, integration, monitoring, and feedback loops in maintaining the operation of ML-based systems in the wild~\cite{makinen2021needs}. We propose incorporating social features in this pipeline by designing for HRO principles such as preoccupation with failure, sensitivity to operations, and deference to expertise. 
For example, include (1) continuous failure monitoring, effectively serving as distributed fire alarms that can be engaged by people at varying levels in an organization, and (2) model maintenance, by relying on people on the ground for detailed understanding of failure cases, as seen in organizations that perform failure panels, audits, etc. 

%% file: 05_conclusion.tex
Interpretability and explainability approaches are designed to help stakeholders adequately understand the predictions and reasoning of an ML-based system. Although these approaches represent complex models in simpler formats, they do not account for the contextual factors that affect whether and how people internalize information. We have presented an alternate framework for helping people understand ML models grounded in Weick's sensemaking theory from organizational studies. Via its seven properties, sensemaking describes the individual, environmental, social, and organizational context that affects human understanding. We translated these for the human-machine context and presented a research agenda based on each property. We also proposed a new framework---Sensible AI---that accounts for these nuances of human cognition and presented initial design ideas as a concrete path forward. We hope that by accounting for these nuances, Sensible AI can support the desiderata (e.g., reliability, robustness, trustworthiness, accountability, fair and ethical decision-making, etc.) that interpretability and explainability are intended for. 